\documentclass{article}
\usepackage{spconf,amsmath,graphicx}
\usepackage{amsfonts}
\usepackage{mathtools,lipsum,cuted}
\usepackage{xfrac}
\usepackage[colorlinks]{hyperref}

\title{Multi-Segment Reconstruction Using Invariant Features}
\name{Mona Zehni, Minh N. Do, Zhizhen Zhao}
\address{Department of ECE and CSL, University of Illinois at Urbana-Champaign}
\begin{document}
%
\maketitle
\begin{abstract}
Multi-segment reconstruction (MSR) problem consists of recovering a signal from noisy segments with unknown positions of the observation windows. One example arises in DNA sequence assembly, which is typically solved by matching short reads to form longer sequences. Instead of trying to locate the segment within the sequence through pair-wise matching, we propose a new approach that uses shift-invariant features to estimate both the underlying signal and the distribution of the positions of the segments. Using the invariant features, we formulate the problem as a constrained nonlinear least-squares. The non-convexity of the problem leads to its sensitivity to the initialization. However, with clean data, we show empirically that for longer segment lengths, random initialization achieves exact recovery. Furthermore, we compare the performance of our approach to the results of expectation maximization and demonstrate that the new approach is robust to noise and computationally more efficient.  
\end{abstract}


\section{Introduction}
\label{sec:intro}

\begin{figure}
\centering
\includegraphics[width=1 \linewidth]{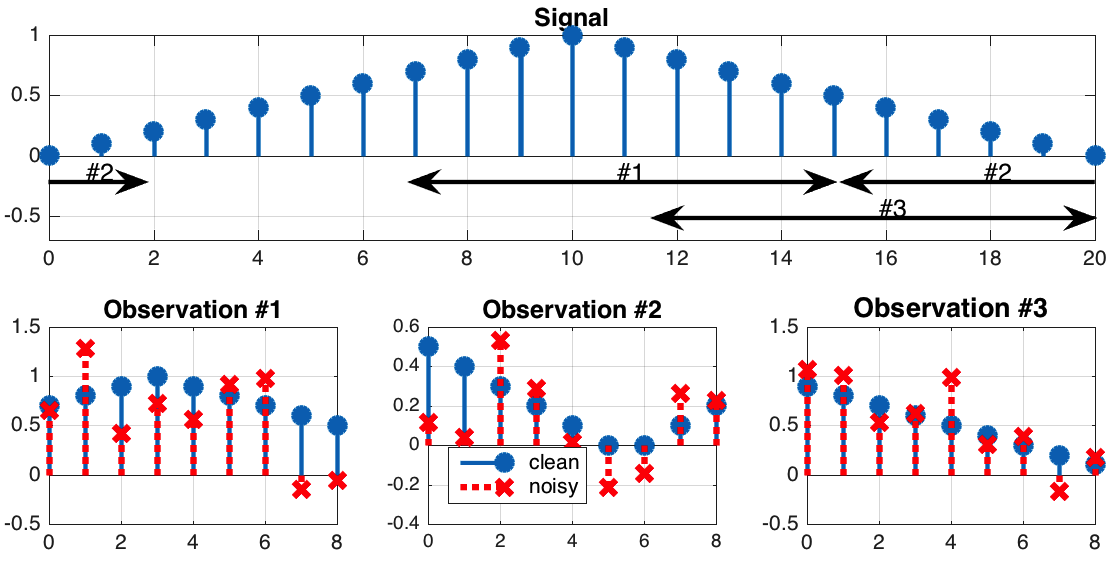}
\caption{(a) The original signal, (b) several noisy circular segments of the signal}
\label{fig:system_model}
\end{figure}

We consider the following observation model,
\begin{equation}
\label{eq:obs_model}
y_k = \mathcal{M}_{s_k} x + \varepsilon_k, \quad k \in \{1,2,...,K\}
\end{equation}
where $y_k \in \mathbb{R}^m$ and $x \in \mathbb{R}^d$, $m \leq d$, correspond to the $k$-th observation and the underlying signal respectively. $\mathcal{M}_{s}$ denotes a cyclic masking operator that captures $m$ consecutive entries of the signal starting from location $s$. In other words, $\mathcal{M}_s: \mathbb{R}^d \rightarrow \mathbb{R}^m$ and $\left(\mathcal{M}_{s}x\right)[n] = x[n+s \, \textrm{mod} \, d]$. For the sake of brief notations, we define $x[n+s]_d \coloneqq x[n+s \, \textrm{mod} \, d]$ from now on. We also assume $s \in \{0,1,...,d-1\}$ to be a random variable drawn from a general distribution with $p$ as its probability mass function, i.e. ${P}\{s=s_k\} = p[s_k]$. 
Furthermore, the randomly located segment of the signal is contaminated by additive white Gaussian noise $\varepsilon_k$ with zero mean and variance $\sigma^2 I_m$. Figure \ref{fig:system_model} further illustrates the observation model \eqref{eq:obs_model}.

Our goal here is to recover $x$ from noisy partial observations $\{y_k\}_{k=1}^K$. This problem is linked to \textit{multi-reference alignment (MRA)} \cite{bandeira2014multireference} in which estimation of the signal from noisy and random circularly-shifted versions of itself is targeted \cite{Bendory2017,Abbe2017}. While in MRA the whole signal takes part in each observation, in MSR shorter segments of the signal contribute to the observations. Similar problems to MSR appear in DNA sequencing \cite{Motahari2013,Green2001}, common superstring problem \cite{frieze1998greedy,kaplan2005greedy,ma2009greed}, puzzle solving \cite{Paikin2015}, image registration, super-resolution imaging \cite{Vandewalle2007} and cryo-EM \cite{Kam1985,Zhao2013}, to name a few.

MSR is originally motivated by DNA sequencing, short common super string (SCS) problem and cryo-EM. While in DNA sequencing, assembling the whole sequence from short-length reads is addressed, in SCS finding the shortest string containing a set of strings is the ultimate target. Also, our signal model is relevant to the cryo-EM 3D reconstruction in the sense that the Fourier transform of a 2D projection image is a partial observation of the 3D volume.

In this paper, we have assumed fixed segment length, however the same approach can be easily extended to random-length segments. In addition, we assume that the probability mass function of the positions of the segments is no longer uniform unlike \cite{Motahari2013}, thus further generalizing the problem. Next, we estimate some features of the signal from the observations and then try to recover the signal from those features. The advantages of pursuing this approach compared to using the observations directly are in \textit{1)} we estimate $x$ and $p$ directly, thus circumventing the estimation of $\mathcal{M}_{s_k}$ which is most of the times impossible due to high level of noise \cite{Barnett2016}-\cite{Shneerson2008}, \textit{2)} creating features that are shift-invariant, i.e. if $x$ and $p$ are shifted the same amount, the constructed features will not change, \textit{3)} instead of using probabilistic models such as maximum-likelihood which are computationally expensive \cite{Barnett2016,Punjani2017,Scheres2016}, our approach goes through the observations once to create invariant features similar to \cite{Bendory2017,Kam1985,Bouman2016}, \textit{4)} the signal recovery is more robust to noise as the features can be estimated accurately when sufficient number of observations is available.

We formulate the problem of recovering $x$ and $p$ as a weighted non-linear least-squares problem. We seek to find a signal that matches the higher order statistics (up to third order correlation) derived from the observations. In addition, due to the structure of the constructed features, the formulated optimization problem can be viewed as a tensor decomposition problem \cite{Kolda2009}-\cite{tensorlab3.0}. Our simulations reveal that this problem has spurious local minima apart from the global minima, hence additional care should be given to the initialization scheme. Note that $x$ and $p$ are only determined up to a global cyclic shift. We can clearly see that as the length of the segment increases, the convergence of the recovery problem to the accurate solution becomes less sensitive to random initialization. Additionally, we observe that it is impossible to recover from partial observations of the signal that are shorter than a threshold, similar to \cite{Motahari2013}. We also apply our approach to some small scale form of gene sequencing problem. Relying on the results we hope we can further extend our problem to larger scales such that it finds real applications in DNA sequence assembly. MATLAB implementations of our paper are provided in https://github.com/MonaZI/MSR.

The organization of the paper is as follows. In Section \ref{sec:methods} we describe our method. In Section \ref{sec:simulation_results} we present the results of our approach and finally conclude the paper in Section \ref{sec:conclusion}.
\section{Methods}
\label{sec:methods}
We use the first, second, and third order correlation of the signal as the shift-invariant features. Let $\mu$, $C$ and $T$ be the population expectations corresponding to these features obtained from clean observations as in \eqref{eq:mean_clean}. 
\begin{align}
\label{eq:mean_clean}
\mu_{x,p}[n]  &= \sum\limits_{s=0}^{d-1} x[n+s]_d\, p[s], \\
C_{x,p}[n_1,n_2]  &= \sum\limits_{s=0}^{d-1} x[n_1+s]_d\, x[n_2+s]_d\, p[s],\nonumber \\
T_{x,p}[n_1,n_2,n_3]  &= \sum\limits_{s=0}^{d-1} x[n_1+s]_d\, x[n_2+s]_d\, x[n_3+s]_d\, p[s]. \nonumber
\end{align}
We use the observations $y_k$ to construct empirical estimates of invariants in \eqref{eq:mean_clean} as in \eqref{eq:mean_true}. \cite{Bendory2017} verifies that the relative error in the estimation of the second and third-order correlation decays as $\frac{1}{\sqrt{K}}$ and the sample complexity for the estimation of $T$ and $C$ is $O({ \sigma^6})$ and $O({\sigma^4 })$ respectively.
\begin{align}
\label{eq:mean_true}
& \widehat{\mu}[n] = \frac{1}{K} \sum\limits_{k=1}^{K} y_k[n] \rightarrow \mu_{x,p}[n], \\
& \widehat{C}[n_1,n_2] =\frac{1}{K} \sum\limits_{k=1}^{K}y_k[n_1]y_k[n_2] - \sigma^2 \delta(n_1,n_2) \rightarrow C_{x,p}[n_1,n_2] \nonumber \\
& \widehat{T}[n_1,n_2,n_3] =\frac{1}{K}\sum\limits_{k=1}^{K}y_k[n_1]y_k[n_2]y_k[n_3] - \sigma^2 \left(\widehat{\mu}[n_1] \delta(n_2,n_3) \right . \nonumber \\
& \left . +\widehat{\mu}[n_2] \delta(n_1,n_3)+ \widehat{\mu}[n_3] \delta(n_1,n_2) \right)  \rightarrow T_{x,p}[n_1,n_2,n_3] \nonumber
\end{align}
Thus, MSR formulation for non-uniform $p$ is described in \eqref{eq:opt_non_unif} where $\Vert . \Vert_F$ marks the Frobenius norm,
\begin{align}
& \min_{x,p} \lambda_T\Vert \widehat{T} - {T}_{x,p}\Vert_F^2 + \lambda_C \Vert \widehat{C} - {C}_{x,p}\Vert_F^2 + \lambda_{\mu} \Vert \widehat{\mu} - {\mu}_{x,p} \Vert_2^2 \nonumber \\
& s.t. \quad {\forall i \in \{0,\ldots,d-1\}, \,}{p[i]\geq 0,\,}\, \sum\limits_{i=0}^{d-1} p[i] =1. \label{eq:opt_non_unif}
\end{align}

In case of uniform distribution for $s$, $p[s=i] = \frac{1}{d}$, $\forall \, i \in \{0,1,...,d-1\}$, the dimension of the invariant features will further reduce due to existing symmetries. As a result, similar derivations to \eqref{eq:mean_clean} simplify as,
\begin{align}
\widetilde{\mu}_{x} = \frac{1}{d} \sum\limits_{m=0}^{d-1} x\left[m\right], \widetilde{C}_{x}\left[n\right] = \frac{1}{d} \sum\limits_{m=0}^{d-1}x[m+n]_d x[m], \nonumber \\
\widetilde{T}_{x}\left[n_1,n_2\right] = \frac{1}{d} \sum\limits_{m=0}^{d-1} x[n_1+m]_d x[n_2+m]_d x[m]. \label{eq:unif_mu}
\end{align}
The MSR formulation for uniform $p$ is,
\begin{align}
\min_{x} \lambda_T\Vert {\widehat{T}} - \widetilde{T}_x\Vert_F^2 + \lambda_C \Vert {\widehat{C}} - \widetilde{C}_x\Vert_F^2 + \lambda_{\mu} \Vert {\widehat{\mu}} - \widetilde{\mu}_x \Vert_2^2 \label{eq:opt_unif}
\end{align}
where we reuse ${\widehat{T}}$, ${\widehat{C}}$ and ${\widehat{\mu}}$ notations to also refer to the empirical estimates of $\widetilde{T}_x$, $\widetilde{C}_x$ and $\widetilde{\mu}_x$ respectively. Although \eqref{eq:opt_unif} is a special case of \eqref{eq:opt_non_unif} there are a few differences that makes the former interesting to study separately. First, the only unknown we seek to recover in \eqref{eq:opt_unif} is the signal $x$ unlike \eqref{eq:opt_non_unif} in which both $x$ and $p$ are undetermined. Also, \eqref{eq:opt_unif} is an unconstrained optimization problem, while \eqref{eq:opt_non_unif} requires $p$ to be a valid discrete probability mass function. Besides, the complexity of \eqref{eq:opt_unif} is further reduced due to lower dimensions of the invariant features. 


Note that the objective functions in both problems  correspond to the weighted squared Frobenius distance between the ground truth features of $x$ and their estimated values. $\lambda_T$, $\lambda_C$ and $\lambda_{\mu}$ are the weights we give to the importance of matching the third, second and first order correlation terms with their estimated values. For example, as $\lambda_T$ increases, $x$ and $p$ are set in a way that further match $\widehat{T}$. In Section \ref{sec:simulation_results}, we also examine the case with $\lambda_T=0$ to see the possibility of accurate recovery of $x$ and $p$ by merely using statistics up to second order.

Note the objective function in \eqref{eq:opt_non_unif} which is a $6$-th order polynomial in $x$ and $2$-nd order polynomial in $p$. Thus, although the constraints form a convex set, the overall optimization problem is non-convex. There are couple of challenges with non-convex optimization, \textit{1)} existence of local minima 
and \textit{2)} existence of saddle points which might slow down the first-order optimization approaches. 
To avoid the second pitfall, we exploit second order methods such as trust-region and sequential quadratic programming (SQP) \cite{Nocedal2006}-\cite{Reddi2017}, implemented in MATLAB optimization toolbox. 

In addition to the local non-convex optimization approach, we use global optimization with polynomials~\cite{Lasserre2001} to reconstruct the signal from the invariant moments. 
More specifically, the objective function is a sum of squares (SOS) polynomial. 
The Lasserre hierarchy of relaxations is able to solve the MSR problem for small $d$ with $m$ above a $d$-dependent threshold\footnote{This will be further discussed in the sequel.}. However, it becomes computationally expensive and requires too much memory for $d>9$. 

\subsection{Analysis}
Here we briefly analyze our problem for the clean case with $\sigma = 0$. It is worth mentioning that as the simultaneous shifts of $x$ and $p$ result in the same features, there are at least $d$ global minima. 
When the segment length is small, the number of algebraically independent equations provided by the invariant features in~\eqref{eq:mean_clean} is not enough to uniquely determine $x$ and $p$. Therefore, random initialization with local non-convex algorithms are able to achieve the global minima, but fail to recover the true signal and the corresponding segment location pmf. 


Let us denote $\tilde{m}(d)$ as the minimum $m$ for which the number of algebraically independent equations provided by the invariant features reaches the number of unknowns. $\tilde{m}(d)$ varies across different problem settings as,
\begin{equation}
\tilde{m}(d) = \min_{m \in \mathbb{N}} m \nonumber
\end{equation}
\begin{equation}
s.t. \quad
\begin{cases}
    \frac{m^3}{6} + m^2 + \frac{11}{6}m + 1 \geq 2d & \textrm{non-unif. } p, \lambda_T \neq 0 \\
  \frac{m^2}{2} + \frac{3}{2} m + 1 \geq 2d & \textrm{non-unif. } p, \lambda_T = 0 \\
  \frac{m^2}{2} + \frac{3}{2} m + 1 \geq d & \textrm{unif. } p, \lambda_T \neq 0
\end{cases}
\label{eq:m_tilde}
\end{equation}

We provide numerical results to verify our analysis of $\tilde{m}(d)$. In addition, we can extend our approach to contain moments up to $s>3$ and count the number of algebraically independent equations in order to determine the minimum required segment length.

In bispectrum inversion for 1D MRA~\cite{Bendory2017}, it is observed that with random initialization, local non-convex algorithm is able to exactly recover the signal in the noiseless case. However, for MSR, $m > \tilde{m}(d)$ is not sufficient to guarantee the exact recovery from random initialization. In some cases, gradient methods can get stuck in the local minima and therefore require good initialization. Similar phenomenon is observed in \cite{Boumal2017} for reconstructing heterogeneous signals from invariant moments. 

\section{Numerical Results}
\label{sec:simulation_results}
\begin{figure*}
\centering
\includegraphics[width = 1 \linewidth]{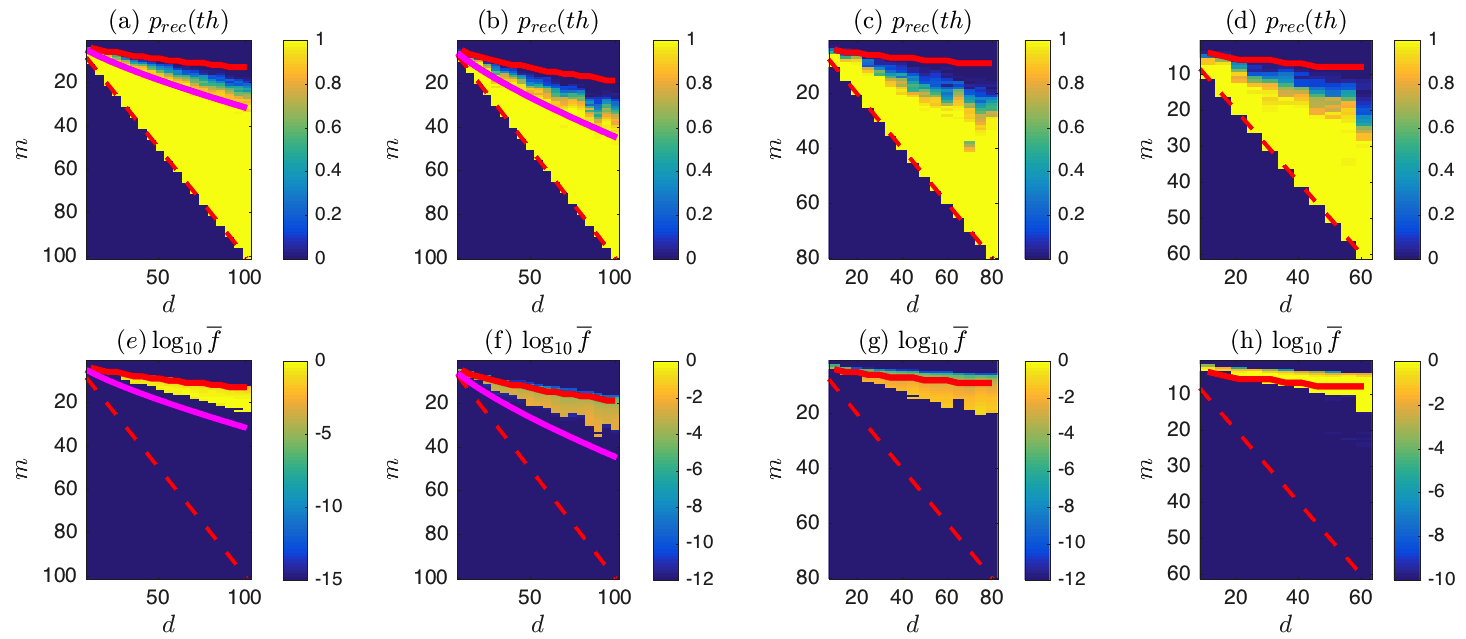}
\caption{$p_{rec}(th)$ and $\overline{f}$ for \textit{a,e)} uniform $p$ and $\lambda_T \neq 0$ in \eqref{eq:opt_unif}, \textit{b,f)} non-uniform $p$ and $\lambda_T = 0$ in \eqref{eq:opt_non_unif}, \textit{c,g)} non-uniform $p$ and $\lambda_T \neq 0$ in \eqref{eq:opt_non_unif}, \textit{d,h)} $x$ has discrete values, non-uniform $p$ and $\lambda_T \neq 0$ in \eqref{eq:opt_non_unif}. The red solid lines mark $\tilde{m}(d)$, the red-dashed lines convey the upper bound on $m$, i.e. $m \leq d$ and the solid magenta line locate the minimum $m$ for each $d$ for which $p_{rec}(th)$ becomes one. For \textit{(a,e)} and \textit{(b,f)} the magenta line is fit to $d^{\frac{3}{4}}$ and $\sqrt{2} d^{\frac{3}{4}}$ respectively.} 
\label{fig:phase_transition}
\vspace{-0.5cm}
\end{figure*}

In our simulations we generate $x$ and $p$ randomly. Also, we adopt $10^5$ noisy observations to estimate the shift-invariant features as in \eqref{eq:mean_true}. Also, in (\ref{eq:opt_non_unif}) and (\ref{eq:opt_unif}) we assume $\lambda_{\mu} = \lambda_C = \lambda_T = 1$ unless otherwise stated. To assess our methodology we use several performance metrics, \textit{1)} mean-squared error defined as $\textrm{MSE} = \Vert x - \hat{x}\Vert^2$, \textit{2)} the probability of accurate recovery, i.e. $p_{rec}(th) = \textrm{P} \{ \textrm{MSE} \leq th \}$, \textit{3)} the median of the final value of the objective function denoted by $\overline{f}$ which is an indicator of whether the globally optimal solution is obtained.
For our evaluations in this section, we set $th = 10^{-3}$. To derive $p_{rec}(th)$ and $\overline{f}$, we solve the optimization problem using trust-region and SQP starting from a random initial point for $50$ trials. Note that when we state accurate recovery is achieved, we mean an accurate estimation of $x$ and $p$ is recovered up to a global cyclic shift and the corresponding MSE is below $th$. In what follows, we discuss the two main results of our experiments.

$\bullet$ \textit{The impact of the segment length on the possibility of getting to the global minima}:
We investigate the changes of $p_{rec}(th)$ and $\overline{f}$ with respect to $m$ and $d$ for four different cases when $\sigma = 0$ as illustrated in Fig. \ref{fig:phase_transition}, \textit{a)} uniform $p$ and $\lambda_T \neq 0$, \textit{b)} non-uniform $p$ and $\lambda_T = 0$, \textit{c)} non-uniform $p$ and $\lambda_T \neq 0$,
\textit{d)} non-uniform $p$, $\lambda_T \neq 0$ and $x$ discretized in value, i.e. $x[n] \in \{0,1,2,3\}$, $\forall n \in \{0,1,...,d-1\}$.

An immediate observation from all four subplots in Fig. \ref{fig:phase_transition} suggests that the larger the $m$, the higher $p_{rec}(th)$. Comparing Fig. \ref{fig:phase_transition}(a) with Fig. \ref{fig:phase_transition}(b) verifies that for uniform $p$, the minimum length of segments required for accurate recovery is smaller compared to non-uniform $p$ and $\lambda_T = 0$ case. Also, for non-uniform $p$ when $\lambda_T \neq 0$, $\tilde{m}(d)$ is smaller compared to the case of $\lambda_T = 0$, as also predicted by \eqref{eq:m_tilde}. This clearly proposes that using the $3$-rd order correlation provides more information about the signal and thus accurate signal recovery can be obtained for smaller $m$. Additionally, Fig. \ref{fig:phase_transition}(d),(h) shows how our proposed method can be extended to problems in which the signal is discretized in value (similar to a DNA sequence assembly problem) and again how accurate recovery is achievable when the length of the reads surpass a certain threshold. 
\begin{figure}[h!]
\centering
\includegraphics[width=0.88 \linewidth]{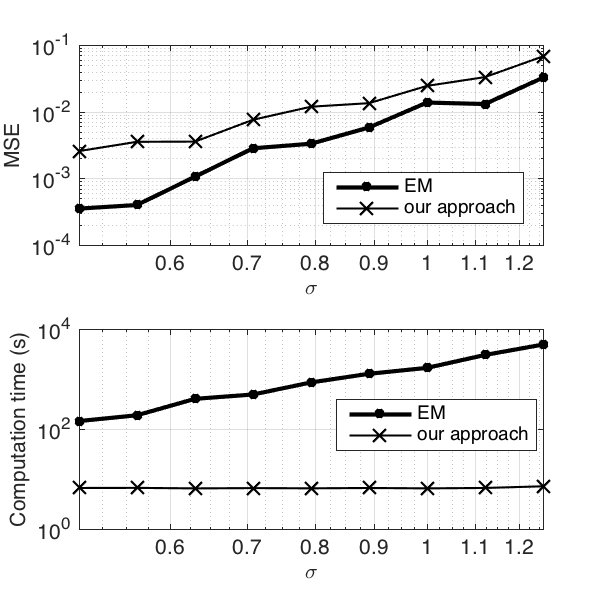}
\vspace{-0.5cm}
\caption{The comparison between the results of our approach and EM in terms of MSE and computation time for different noise levels and fixed $d=45$ and $m=25$.}
\label{fig:ML_cov}
\vspace{-0.5cm}
\end{figure}

Regarding the landscape of the problem what we observe is, \textit{1)} for small values of $m$ the global minimum is not unique (up to cyclic shifts) and reaching global minima does not necessarily guarantee accurate reconstruction and, \textit{2)} the problem has local minima. The evidence for the first statement is that for some trials, although the value of the objective function at the optimal point is reported very small ($\sim 10^{-10}$), $\hat{x}$  and $\hat{p}$, do not match their true values, as displayed in Fig. \ref{fig:phase_transition}. The $d-m$ region for which this happens is the blue-colored region on top of the yellow strip in Fig. \ref{fig:phase_transition}(e)-(h) which almost maps to $m < \tilde{m}(d)$ region. Additionally, we noticed that in some trials, when reaching the local minimum is reported with relatively larger values of the objective function at the optimal point, the solution does not match the original $x$ and $p$. This also marks the existence of local minima in addition to global minima. The corresponding $d-m$ region for this case is marked by the yellow shaded regions in Fig. \ref{fig:phase_transition}(e)-(h).

$\bullet$ \textit{Robustness of the recovery to noise and comparison with expectation-maximization (EM) method}:
Figure \ref{fig:ML_cov} compares the performance of our approach with the results obtained from expectation maximization \cite{dempster1977maximum} for different noise levels. It can be inferred that in high noise regimes the performance of both our approach and EM degrades. On the other hand, our approach is computationally more efficient and scales linearly with the number of samples, so it can be used as a good initialization for EM.

\section{Conclusion}
\label{sec:conclusion}

In this paper, we proposed a new approach for recovering a signal from a large number of randomly observed noisy segments. The random locations of the observation windows are unknown. Instead of trying to recover the locations for each segment through matching, we used shift invariant features to estimate the underlying signal and the distribution of the windows. The invariant features approach has low computational complexity for large sample size compared to alternative methods, such as EM.  The signal is reconstructed by solving a constrained nonlinear least-squares problem. Due to the non-convex nature of the problem, the solution depends on the initialization. It was shown that for clean data, as the length of the segment increases, random initialization can achieve accurate recovery. We also demonstrated that the new method is robust to noise and efficient in terms of computational time. 

\bibliographystyle{IEEEbib}
\bibliography{ADMMref.bib}

\begin{thebibliography}{10}

\bibitem{bandeira2014multireference}
A.~S. Bandeira, M.~Charikar, A.~Singer, and A.~Zhu,
\newblock ``Multireference alignment using semidefinite programming,''
\newblock in {\em Proceedings of the 5th conference on Innovations in
  theoretical computer science}. ACM, 2014, pp. 459--470.

\bibitem{Bendory2017}
T.~{Bendory}, N.~{Boumal}, C.~{Ma}, Z.~{Zhao}, and A.~{Singer},
\newblock ``{Bispectrum Inversion with Application to Multireference
  Alignment},''
\newblock {\em IEEE Transactions on Signal Processing}, vol. 66, no. 4, pp.
  1037--1050, 2017.

\bibitem{Abbe2017}
E.~{Abbe}, T.~{Bendory}, W.~{Leeb}, J.~{Pereira}, N.~{Sharon}, and A.~{Singer},
\newblock ``Multireference alignment is easier with an aperiodic translation
  distribution,''
\newblock {\em arXiv preprint arXiv:1710.02793}, Oct. 2017.

\bibitem{Motahari2013}
A.~S. Motahari, G.~Bresler, and D.~N.~C. Tse,
\newblock ``Information theory of {DNA} shotgun sequencing,''
\newblock {\em IEEE Transactions on Information Theory}, vol. 59, pp. 6273 --
  6289, 2013.

\bibitem{Green2001}
E.~D. Green,
\newblock ``Strategies for the systematic sequencing of complex genomes,''
\newblock {\em Nature Reviews, GENETICS}, vol. 2, pp. 573--583, 2001.

\bibitem{frieze1998greedy}
A.~Frieze and W.~Szpankowski,
\newblock ``Greedy algorithms for the shortest common superstring that are
  asymptotically optimal,''
\newblock {\em Algorithmica}, vol. 21, no. 1, pp. 21--36, 1998.

\bibitem{kaplan2005greedy}
Haim Kaplan and Nira Shafrir,
\newblock ``The greedy algorithm for shortest superstrings,''
\newblock {\em Information Processing Letters}, vol. 93, no. 1, pp. 13--17,
  2005.

\bibitem{ma2009greed}
Bin Ma,
\newblock ``Why greed works for shortest common superstring problem,''
\newblock {\em Theoretical Computer Science}, vol. 410, no. 51, pp. 5374--5381,
  2009.

\bibitem{Paikin2015}
G.~Paikin and A.~Tal,
\newblock ``Solving multiple square jigsaw puzzles with missing pieces,''
\newblock in {\em Computer Vision and Pattern Recognition (CVPR), 2015 IEEE
  Conference on}, 2015, pp. 161--174.

\bibitem{Vandewalle2007}
P.~Vandewalle, L.~Sbaiz, J.~Vandewalle, and M.~Vetterli,
\newblock ``Super-resolution from unregistered and totally aliased signals
  using subspace methods,''
\newblock {\em IEEE Transactions on Signal Processing}, vol. 55, pp. 3687 --
  3703, 2007.

\bibitem{Kam1985}
Z.~Kam and I.~Gafni,
\newblock ``Three-dimensional reconstruction of the shape of human wart virus
  using spatial correlations,''
\newblock {\em Ultramicroscopy}, vol. 17, pp. 251--262, 1985.

\bibitem{Zhao2013}
Z.~{Zhao} and A.~{Singer},
\newblock ``Rotationally invariant image representation for viewing direction
  classification in cryo-{EM},''
\newblock {\em Journal of structural biology}, vol. 186, no. 1, pp. 153--166,
  2014.

\bibitem{Barnett2016}
Alex Barnett, Leslie Greengard, Andras Pataki, and Marina Spivak,
\newblock ``Rapid solution of the cryo-em reconstruction problem by frequency
  marching,''
\newblock {\em SIAM Journal on Imaging Sciences}, vol. 10, no. 3, pp.
  1170--1195, 2017.

\bibitem{Shneerson2008}
V.~L. Shneerson, A.~Ourmazd, and D.~K. Saldin,
\newblock ``Crystallography without crystals. {I}. the common-line method for
  assembling a three-dimensional diffraction volume from single-particle
  scattering,''
\newblock {\em Acta Crystallographica}, 2008.

\bibitem{Punjani2017}
A.~Punjani, M.~A. Brubaker, and D.~J. Fleet,
\newblock ``Building proteins in a day: Efficient 3{D} molecular structure
  estimation with electron cryomicroscopy,''
\newblock {\em IEEE Transactions on Pattern Analysis and Machine Intelligence},
  vol. 39, pp. 706--718, 2017.

\bibitem{Scheres2016}
S.H.W. Scheres,
\newblock ``Chapter {S}ix - {P}rocessing of structurally heterogeneous
  cryo-{EM} data in {RELION},''
\newblock {\em Methods in Enzymology}, vol. 579, pp. 125--157, 2016.

\bibitem{Bouman2016}
K.~L. {Bouman}, M.~D. {Johnson}, D.~{Zoran}, V.~L. {Fish}, S.~S. {Doeleman},
  and W.~T. {Freeman},
\newblock ``{Computational Imaging for VLBI Image Reconstruction},''
\newblock in {\em The IEEE Conference on Computer Vision and Pattern
  Recognition (CVPR)}, Mar. 2016.

\bibitem{Kolda2009}
T.~G. Kolda and B.~W. Bader,
\newblock ``Tensor decompositions and applications,''
\newblock {\em SIAM REVIEW}, vol. 51, no. 3, pp. 455--500, 2009.

\bibitem{tensorlab3.0}
N.~Vervliet, O.~Debals, L.~Sorber, M.~Van~Barel, and L.~De~Lathauwer,
\newblock ``Tensorlab 3.0,'' Mar. 2016,
\newblock Available online.

\bibitem{Nocedal2006}
J.~Nocedal and S.~J. Wright,
\newblock {\em Numerical Optimization},
\newblock Springer Series in Operations Research and Financial Engineering,
  2006.

\bibitem{Reddi2017}
S.~J {Reddi}, M.~{Zaheer}, S.~{Sra}, B.~{Poczos}, F.~{Bach},
  R.~{Salakhutdinov}, and A.~J {Smola},
\newblock ``{A Generic Approach for Escaping Saddle points},''
\newblock {\em arXiv preprint arXiv:1709.01434}, Sept. 2017.

\bibitem{Lasserre2001}
Jean~B. Lasserre,
\newblock ``Global optimization with polynomials and the problem of moments,''
\newblock {\em SIAM Journal on Optimization}, vol. 11, no. 3, pp. 796--817,
  2001.

\bibitem{Boumal2017}
N.~{Boumal}, T.~{Bendory}, R.~R. {Lederman}, and A.~{Singer},
\newblock ``{Heterogeneous multireference alignment: a single pass approach},''
\newblock {\em arXiv preprint arXiv:1710.02590}, Oct. 2017.

\bibitem{dempster1977maximum}
Arthur~P Dempster, Nan~M Laird, and Donald~B Rubin,
\newblock ``Maximum likelihood from incomplete data via the {EM} algorithm,''
\newblock {\em Journal of the royal statistical society. Series B
  (methodological)}, pp. 1--38, 1977.

\end{thebibliography}

\end{document}